\documentclass[twocolumn]{aastex62}
\usepackage{placeins} % remove when draft is ready.
\usepackage{natbib}
\usepackage{apjfonts} 
\usepackage{xcolor}
\usepackage{multirow}
\newcommand{\msun}{M$_{\sun}$}

\newcommand{\Teff}{$T_{eff}$}
\newcommand{\lgTeff}{$log_{10}(T_{eff}/\rm K)$}
\newcommand{\lgL}{$log_{10}(L/L_{\odot})$}
\newcommand{\kmps}{$\rm km \ s^{-1}$}

\newcommand{\powten}[1]{$\rm 10^{#1}$}

\received{XXX}
\revised{YYY}
\accepted{ZZZ}
\shorttitle{B/R ratio for supernova progenitor models at low metallicities}
\shortauthors{Wagle et al.}
\begin{document}

%\title{Convective Overshoot In Type II-P Supernova Progenitors: Surface Composition, Blue Loops And Their Explodability}
\title{Type IIP Supernova Progenitors III: Blue to Red Supergiant Ratio in Low Metallicity Models with Convective Overshoot}

\correspondingauthor{Gururaj A. Wagle}
\email{guru.w84@gmail.com}

\author[0000-0002-3356-5855]{Gururaj A. Wagle}
\affiliation{Homi Bhabha Centre for Science Education - Tata Institute of Fundamental Research, Mankhurd, Mumbai 400088, India}

\author[0000-0003-2404-0018]{Alak Ray}
\affiliation{Homi Bhabha Centre for Science Education - Tata Institute of Fundamental Research, Mankhurd, Mumbai 400088, India}

\author{Adarsh Raghu}
\altaffiliation{Participant in the National Initiative on Undergraduate Science (NIUS)\\ program
at HBCSE (TIFR)}
\affiliation{Indian Institute of Science Education and Research Kolkata, 741246, India}


\begin{abstract}

The distribution of stars in the Hertzsprung Russell diagram (HRD) for a stellar conglomeration represents a snapshot of its evolving stellar population. Some of the supergiant stars may transit the HRD from blue to red and then again to blue during their late evolutionary stages, as exemplified by the progenitor of SN 1987A. Others may transit a given part of the HRD more than twice in a ``blue loop'' and end up as red supergiants before they explode\deleted{ as in Type IIP SN 2013ej}. Since stars in blue loops spend a considerable part of their lives there, these stages may change the relative number of modeled supergiants in the HRD. Their lifetimes in turn depend upon the initial mass of the star, how convection in its interior is modeled, and how much mass loss takes place during its evolution. The observed ratio of the number of blue to red supergiants and yellow to red supergiants sensitively test the stellar evolution theory. We compare modeled number ratios of these supergiants with observed data from the Large Magellanic Cloud as it has a metallicity very similar to that of the environment of SN 2013ej. We successfully model these by taking into account moderate (exponential) convective overshooting. We explore its effect on the final radius and mass of the star prior to core collapse. The radius differs dramatically with overshoot. These factors controlling pre-supernova structure may affect the post-explosion optical/IR light curves and spectral development.

\end{abstract}



\section{Introduction} \label{sec:intro}

Massive stars transit red-ward on the Hertzsprung Russell diagram (HRD) from a blue supergiant (BSG) phase after they have fully exhausted hydrogen in their core. The BSG (B0 to A9 spectral types) progenitor enters the red supergiant (RSG - K0 to M3 I type) phase after passing through a yellow supergiant (YSG - F0 to G9 I) phase. Some of the RSGs may evolve back towards blue at an intermediate stage of their lives, passing through the YSG phase for a second time. These stars at the BSG phase after it has already been a RSG may collapse to give rise to a supernova like SN 1987A \citep{Arnett:1989aa}. Other stars, after traversing a part of their blue-ward track, may turn back and end up in the RSG stage before they collapse to give rise to the more canonical type IIP or IIL supernovae. Many of these RSGs retain a lot of hydrogen envelope at the time of collapse. This doubling-back of the track of the supergiants in the HRD is known as the blue loop. Stars having blue loops spend a significant part of their lifetimes during core helium burning in the blue part of the HRD. The existence of blue loops in the evolutionary tracks may change the number distribution of different types of supergiants in the HRD.

The distribution of stars in the HRD of a system like the LMC, where all stars are at approximately the same distance from us gives us a snapshot of the stellar conglomeration in transition over its very long life. The number of stars in a given region while they are in transit from one part of the HRD to another in a steady state is proportional to the lifetime of that given phase in the HRD. Thus, this distribution reveals how massive stars are actually evolving over time. For a comparison between the models and the observed data, we need to know how many stars are in the system for a given mass range (the initial mass function), since stellar lifetimes are known to vary strongly with the initial mass of the stars. Another important aspect of the stellar lifetimes is how convection goes on in the stellar interior, both in the nuclear fuel burning core regions, as well as, in the convective envelope which forms the star's outer regions. The stellar evolution code MESA models the extra mixing of elements beyond the convective boundary by "convective overshooting", which is parameterized in terms of a scale factor $f$ that operates on the local pressure scale height. The luminosity of the star at any given time is not only controlled by its initial mass but also by the extent of the convective overshoot. 

The intermediate stage YSGs are relatively rare \citep[see][]{Drout:2009aa,Neugent:2012aa,Neugent:2010aa}. A sample of these supergiants that is complete can provide a ``magnifying glass" \citep{Kippenhahn:1990aa} for tests of stellar evolution theory in a particular environment. %Among the various tests that stellar evolutionary theory can be subjected to the ratio of the numbers of YSGs to RSGs (Y/R) and that of the number of BSGs to RSGs (B/R) offer sensitive tests, and for many realisations of stellar evolutionary tracks, theory most often fails to match observed data. As only some computational realisations depending upon the parameters used for evolutionary calculations can match this and other observational results, these tests can collectively offer significant insights for limiting the parameter space of stellar evolutionary models that the imperfect modelling\footnote{due to the limitations of computational resources to model real stars} of the stars' evolution can undertake. 
Moreover, as \citet{Neugent:2010aa} point out, it is vital to have reliable evolutionary tracks of the luminous stars to interpret the spectra of distant galaxies using population synthesis codes as often the individual stars cannot be resolved in these galaxies. Stellar evolutionary models are also necessary to study mixed age populations of stars and their initial mass functions.

Computing the ratio of numbers of blue to red supergiants (B/R) and of yellow to red supergiants (Y/R) to match the data in varying galactic environments is thus one of the most challenging problems in astrophysics. The B/R ratio has been observed to be a steeply rising function of metallicity, Z in an ensemble of massive stars in galaxies or a set of young clusters \citep[see][and references therein]{Langer:1995aa}. However, most of the model calculations fail to fit the ratio for a given set of parameters as a function of metallicity. 
The ratio is sensitive to mass loss, convection and other mixing processes\added{, such as due to stellar rotation induced mixing \citep{Maeder:2001aa}}. Since convective overshoot is an important ingredient of mixing in both the nuclear burning core or its edge as well as in the stellar envelope surrounding it, the B/R ratio 
%\added{in a restricted region of the Hertzsprung Russell diagram (HRD)} 
may be an important diagnostic of the extent of convective overshoot as well. In this paper, we explore the effects of convective overshoot on the B/R ratio in the case of low metallicity models (Z = 0.006). In our companion Paper I \citep{Wagle:2019aa}, we have discussed how the convective overshoot affects the evolution of a 13 \msun \ star (with Z = 0.006) in terms of its internal structure as well as the evolution in the HRD, especially the existence of blue loop. Here, we extend the methods to a range of stellar masses (12--15 \msun) to study the effects of convective overshoot on the predicted B/R ratio. We then compare our predicted values to the observed ratio for the Large Magellanic Cloud (LMC), which has similar metallicity value (Z = 0.007) and sufficient recent observational data to make this comparison viable. 

In section \ref{sec:obs_BR}, we review the literature for the observed B/R ratio as a function of metallicty in the LMC and other galaxies. In section \ref{sec:obs_HR} we review the selection of observed LMC supergiants for comparing a subset of the data from \citet{Neugent:2012aa} with our model evolutionary calculations. In section \ref{sec:methods}, we discuss the computational methods for stellar evolution models using the code Modules for Experiments in Stellar Astrophysics \citep[MESA,][]{Paxton:2011aa,Paxton:2013aa,Paxton:2015aa,Paxton:2018aa}. In section \ref{sec:results}, we describe the calculations of predicted B/R and Y/R ratios using the models discussed in the previous section. In addition, we discuss the observed B/R and Y/R ratios for the LMC to compare with the predicted ratios. We also discuss the variation of the final radius and mass of the pre-supernova star with overshoot factor. In section \ref{sec:conclusion}, we mention how these variations may affect the post-explosion supernova light curves and spectra along with our conclusions.

\section{B/R Ratio and Its Dependence on Metallicity} \label{sec:obs_BR}

The ratio of B/R supergiants is known to be particularly sensitive to the metallicity of the environment where the supergiant stars reside. The ratio and its Z-dependence obtained from a number of clusters in the Milky Way galaxy and the Magellanic Clouds have been re-examined by \citet{Eggenberger:2002aa}. They give a normalized relation of the B/R (to the solar neighborhood value) in terms of metallicity as:
\begin{equation} \label{eqn:BR_Eggen}
{(B/R) / (B/R)_{\odot}} \sim 0.05 \times e^{{3 Z/Z_{\odot}}}
\end{equation}
\citet{Eggenberger:2002aa} find that the ratio at solar metallicity $(B/R)_{\odot} =$ 3 when including O, B and A supergiants in blue for \textit{log age} interval of 6.8--7.5. They chose stars in stellar clusters instead of field stars as the stars in the clusters have same distance, age and chemical composition. They identified the supergiants based on spectroscopic measurements instead of photometric colors to accurately differentiate supergiants and main sequence stars. They showed that the B/R ratio remains more or less the same even if different \textit{log age} intervals are chosen. They however do not give a value for B/R ratio of the LMC due to lack of sufficient spectroscopic data for the young clusters in the LMC. If we consider the relation given above in equation \ref{eqn:BR_Eggen}, then B/R ratio for the LMC would be around 0.4, assuming a ratio 3.6 for solar neighborhood (counting only the B-type stars). \citet{Eggenberger:2002aa} also tabulate the ratio for the photometric count with limiting visual magnitude ($M_{vlim}$) in their table 3. This value varies from 2.5 at $M_{vlim} = -$3.25 to 5.7 at $M_{vlim} = -$2.0 for the LMC.
\citet{Langer:1995aa} note that the observed B/R ratio in literature is found to be around 0.6 when strictly B and red supergiants are counted for the young cluster NGC 330 in the SMC. However, this value is based on the observations of a single star cluster. They found the same ratio to be 3.6 for young clusters in the solar neighborhood.  \citet{Humphreys:1984aa}  get the ratio of B/R of 10 for the LMC \& 28 for the solar neighborhood, if all O, B and A stars are included in blue. However, these blue counts include main-sequence stars as well. %in \citet{Humphreys:1984aa} data.
\citet{Langer:1995aa} give B/R ratio of 10 for stars and associations in the LMC for limiting bolometric magnitude, $M_{bol} = -$7.5.  However, as mentioned earlier these blue counts include main sequence stars.

\citet{Humphreys:1979aa} note that the B/R ratio determined as a function of luminosity decreases with decreasing luminosity. They also find that the B/R ratio changes less strongly with galactocentric distance (and hence, metallicity) if the ratio is restricted to less luminous supergiants $M_{bol} \ge -$8.5. 

The calculated B/R ratio is extremely sensitive to model parameters like metallicity, mass loss, convection and other mixing processes \citep[see][]{Maeder:2001aa,Langer:1995aa}.  For a given set of input parameters, the computational models have trouble predicting the B/R ratio for both the high and the low end of the metallicities, simultaneously. There are more number of RSG in the SMC (Z = 0.002) than the models have previously predicted. \citet{Langer:1995aa} have reviewed previous work that test the B/R ratios in different environments. The models that do not use \added{convective} overshooting or semiconvection could realize acceptable agreement for the solar neighborhood but failed to produce the relative number of observed red supergiants \deleted{for observed} at low Z \citep[e.g.][]{Brunish:1986aa,Brunish:1982aa,Brunish:1982ab}. \citet{Stothers:1992aa} addressed this problem and showed that use of Ledoux criterion \added{for convection} reproduce\added{s} well the observed B/R ratio at low Z environment of the SMC, but not for Z $>$ 0.006 (e.g. for the LMC and the solar neighborhood). Further, models by \citet{Arnett:1991aa} that use Ledoux criterion and semiconvection fit well the distribution of supergiants in the LMC as given in \citet{Fitzpatrick:1990aa}. \citet{Langer:1995aa} summarize that the models with Schwarzschild's criterion (with or without convective overshoot mixing) are favorable for solar metallicity environments, while \replaced{convection}{models} using the Ledoux criterion with presence of overshooting at the core as well as the base of hydrogen envelope, and \replaced{inclusion}{including} semiconvection yield better results at low Z. 

\section{Selection of the Observed LMC Supergiants for Model Comparison} \label{sec:obs_HR}

%%%%%%%%% TABLE %%%%%%%%%%%
\begin{deluxetable*}{ccccc}
\tablecaption{The observed LMC supergiants and selection criteria \label{tab:supergiants}}
%\tablewidth{0pt}
%\tabletypesize{\scriptsize}
\tablehead{
\colhead{Selection} &  
\multicolumn{4}{c}{Observed Numbers in the data for} 
\\ [-2ex] 
\colhead{Criteria} & \colhead{BSG} &
\colhead{YSG} & \colhead{RSG} & \colhead{Total Supergiants}
}
%\colnumbers
\startdata
\citet{Neugent:2012aa} & \multicolumn{2}{c}{------------ 1528 ------------} & 865 & 2393 \\
\hline
Associated Vizier Catalog & \multicolumn{2}{c}{------------ 1447 ------------} & 521 & 1968 \\
\hline
Category 1 & \multicolumn{2}{c}{------------\phn 309 ------------} & 506 & \phn 815 \\
\hline
\multirow{2}{*}{Effective Temperature} & (\lgTeff $ >$ 3.875) & (3.875 $\ge $ \lgTeff $ >$ 3.68) & (3.68 $\ge $ \lgTeff ) \\
 & 163 & 109 & 543 & \phn 815 \\
 \hline
\multirow{2}{*}{Effective Temperature} & (4.0 $\ge $ \lgTeff $ >$ 3.875) & (3.875 $\ge $ \lgTeff $ >$ 3.68) & (3.68 $\ge $ \lgTeff ) \\
 & 112 & 109 & 543 & \phn 764 \\
\hline
4.0 $<$ \lgL $<$ 5.0 & \phn 97 & \phn 87 & 514 & \phn 698 \\
\hline
4.2 $<$ \lgL $<$ 5.0 & \phn 97 & \phn 87 & 430 & \phn 614
\enddata
\tablecomments{The LMC supergiants observed by \citet{Neugent:2012aa} are listed in this table along with various criteria used to limit the observation sample in order to make comparison with the model calculations of the B/R and Y/R ratios. \citeauthor{Neugent:2012aa} initially observed 2393 supergiants (first row), however, they found nothing but sky \added{(see text in section \ref{sec:obs_HR} for more details)} for \replaced{about 6\%}{69} YSGs and \replaced{60\%}{343} RSG candidates. Most of these missing RSG candidates are hypothesized to be dust-enshrouded objects. The associated Vizier Catalog \citep{Neugent:2012ab} contains candidates excluding those that yielded nothing but the sky in spectroscopic observations. Category 1 label is used to identify the LMC membership. We re-distribute the observed category 1 stars in the BSG, YSG and RSG categories using the effective temperature criteria. A more restrictive \Teff $\le$ 10,000 K criteria is used for BSG candidates, as \citeauthor{Neugent:2012aa} claim that their data is complete up to \Teff $\approx$ 10,000 K. The luminosity limits are applied to select candidates for comparison with the models. The numbers in the last row are used for comparison with the model calculations.}
\end{deluxetable*}
%%%%%%%%%%%%%%%

%\textcolor{red}{\textbf{Note to the referee: The text appearing in orange in this section has been moved from its original location in the erstwhile submitted manuscript to either avoid repetition or for clarity. The same text is marked in its original location as deleted and notes are added next to it to indicate where it has been moved. The newly added text is displayed in magenta color as before.}}

There are many surveys of the LMC supergiants available in the literature. However, most of the observational studies focus on the red and blue supergiant populations, separately. For example, \citet{Davies:2018ab} lists 225 previously observed RSGs+YSGs in the LMC, out of which 200 RSGs have determined spectral types. On the other hand, \citet{Fitzpatrick:1990aa} only consider the LMC stars with spectral type between O3 and G7. \citet{Urbaneja:2017aa} also observed spectra of 32 LMC blue supergiants to study gravity-luminosity relationship. Different surveys introduce different methodology of gathering and analyzing the data which may introduce different biases in each survey. It may be difficult to account for these differences and determine the correction factors of incompleteness, etc. Therefore, we need a complete dataset that includes observations of both blue and red supergiant stars observed in the same field(s) or cluster(s) and treats data analysis and census in a homogeneous way to be able to determine the relevant observed B/R and Y/R ratios. The dataset published by \citet{Neugent:2012aa} comes closest to this approximation. They observed supergiant stars in the LMC using the Cerro Tololo 4 m telescope and a 138-fiber multi-fiber, multi-object spectrometer \textit{Hydra} in 64 fields (each of which has a 40 \textit{arcmin} field of view). They include both RSGs (K and M type\footnote{See a discussion of spectral types, temperatures, luminosity classes of cool supergiants in the LMC by \citet{Dorda:2016aa}} supergiants) and YSGs (F and G-type supergiants) as well as some BSG stars (B and A type\footnote{See the discussion by \citet{Urbaneja:2017aa} for types and temperature range of LMC BSGs.} supergiants). 

\added{In the following subsections \ref{subsec:neugents_criteria} and \ref{subsec:LMC_membership}, we summarize the candidate selection for observations and determination of the LMC membership of the observed candidates by \citet[][see their sections 2 and 3 for further details]{Neugent:2012aa}. In the subsection \ref{subsec:our_criteria}, we define the temperature and luminosity criteria imposed by us to further shortlist the candidates from \citeauthor{Neugent:2012aa} dataset for comparison with our models. In the last subsection \ref{subsec:dust-enshrouded_RSG}, we discuss the dust enshrouded RSGs that were targeted by \citeauthor{Neugent:2012aa} but not detected by their comparatively sensitive observations.}

\subsection{Candidate selection for observations by \citeauthor{Neugent:2012aa}}\label{subsec:neugents_criteria}

\citet{Neugent:2012aa} identified the LMC YSG\footnote{We note here a difference in the nomenclature between the work of \citeauthor{Neugent:2012aa} and ourselves. The former authors' YSG dataset is complete up to the effective temperature of \lgTeff $\approx$  4.0. For LMC, these temperatures include the stars which are traditionally B- or A-type supergiants apart from the F- and G-type supergiants that are traditionally deemed ``yellow'' in the associated Vizier catalog. We thus subdivide their ``YSGs'' into BSG phase if \lgTeff  $>$ 3.875, and in the YSG phase if 3.875 $\ge $ \lgTeff  $>$ 3.68. We include their ``YSGs'' that have \lgTeff  $<$ 3.68 in to the RSG counts.}  and RSG candidates in a multi-step process. They initially selected the \added{probable} YSG and RSG candidates from the US Naval Observatory CCD Astrograph Catalog Part 3 (UCAC3), within a 3.5 degree radius window centered on the LMC's visible disk and with data on color and magnitude ranges \added{obtained from 2 Micron All Sky Survey (2MASS) \citep{Skrutskie:2006aa}}\deleted{, based on stars}. They used UCAC3 quality codes to eliminate galaxies, clusters and double stars in the field. \added{Thus this restricts the stellar sample to single stars as binaries would be discarded by the selection quality codes.}

\replaced{The authors}{\citeauthor{Neugent:2012aa}} claimed that the following procedure gave a complete sample of the YSGs down to 12 \msun \ \added{based on their models}. In selecting the color magnitude ranges they used a \Teff \ range of 4800--7500 K for the YSGs and defined a K magnitude limit as a function of J$-$K for a 12 \msun \ star obtained from ATLAS9 atmosphere models \citep{Kurucz:1992aa}. They, however, defined a flat K-magnitude cutoff at $K_{2MASS} =10.2$ mag when $(J-K)_{2MASS} >  0.9$ for the RSGs \citep[see figure 2 of][]{Neugent:2012aa}. Using these criteria they obtained 2187 \added{probable} YSG candidates and 1949 \added{probable} RSG candidates from the UCAC3 catalog.

\citet{Neugent:2012aa} could however observe only 1528 \added{unique} YSG candidates and 865 \added{unique} RSG candidates. These observed numbers are a fraction (70\% for YSG and 44\% for the RSGs) of the larger number of candidates selected from the UCAC3 catalog mentioned above. They \added{further} discovered that \replaced{4\%}{69} of the \added{1528} YSG candidates and \replaced{about 59\%}{343} of \added{the 865} RSG candidates \added{that they observed} yielded nothing but sky \replaced{upon examining the spectra}{background when their spectra at these sky coordinates were examined}. \added{We will revisit these ``invisible" candidates in subsection \ref{subsec:dust-enshrouded_RSG}.}

The \added{finally} published Vizier catalog \added{\citep{Neugent:2012ab}} \replaced{has}{consists of} 1447 YSG candidates and 521 RSG candidates. This is 95\% of the YSGs and 60\% of the RSG candidates initially observed by \citet{Neugent:2012aa}.
%Their listing of the unique YSG and RSG {\it candidates} in the LMC are smaller than the subsample of the UCAC catalog because their 64 fields covered less sky than that of the %UCAC3 catalog

\subsection{LMC Membership}\label{subsec:LMC_membership}
 %based on their radial velocities, which were clearly distinguished from the foreground stars. They labeled the stars with radial velocity higher than 200 \kmps \ as ``category 1'' -- the probable LMC supergiants. We selected these stars for the comparison between our models and the observations. They claim that their data is complete down to 12 \msun \ based on the J \& K magnitudes calculated from their Geneva models.
 
A reliable method for the removal of the foreground (dwarf) star contamination for the \replaced{stellar}{supergiant} samples in the LMC and other nearby galaxies has been a significant concern for many studies. It is straightforward to separate the extragalactic RSGs from the foreground red dwarfs and giants by a two color diagram \citep{Massey:1998aa,Massey:2009aa}. However, this method does not work well for the YSGs \citep[see][]{Drout:2009aa}, since there is little or no separation of tracks in the two-color diagram for \added{dwarfs and the} late F-type through early K-type supergiants. %For YSGs in M31  
This is \replaced{however}{thus} more of a problem for the YSGs in the LMC than for the RSGs. Foreground stars contaminating the field that contains the target stars of the LMC were weeded out by excluding those stars with absolute proper-motion values greater than  15 $mas \; yr^{-1}$ in RA or Dec. In fact, \citet{Neugent:2012aa} %showed that they
successfully identified the RSGs in the LMC using color and proper motion alone. For the YSGs they required the additional criterion of {\it radial velocities} to separate these from the galactic foreground stars which have distinctly separate and lower radial velocity distribution. The large systemic radial velocity of the LMC and its clear separation from the velocity distribution of the stars in the Milky Way is a key factor in determining the LMC membership. \deleted{Nevertheless, since the blue/yellow to red ratio of supergiants in the LMC has been debated in the literature over a long time and is critical for comparing with model comparisons, we recapitulate below briefly the multiple filters by which the number of observed supergiants are identified and selected.}% {\textbf{Note for the referee: The preceding deleted text has been moved to section \ref{subsec:our_criteria}, and is shown in orange.}}

\deleted{The number of observed BSG, YSG and RSGs arising from various selection criteria relevant to our model calculations are summarized in our Table \ref{tab:supergiants}.} %{\textbf{Note for the referee: The preceding deleted text has been moved to section \ref{subsec:our_criteria}, and is shown in orange.}} 
\deleted{\citet{Neugent:2012aa} selected candidates listed the UCAC3 catalog based on color and magnitude ranges for YSGs and RSGs. The authors claimed that this procedure gave a complete sample of the YSGs down to 12 \msun . In selecting the color magnitude ranges they used a \Teff \ range of 4800--7500 K for the YSGs and defined a K magnitude limit as a function of J$-$K (see their figure 2) for a 12 \msun \ star obtained from ATLAS9 atmosphere models \citep{Kurucz:1992aa}. They, however, defined a flat K-magnitude cutoff at $K_{2MASS} =10.2$ mag when $(J-K)_{2MASS} >  0.9$ for the RSGs, obtaining 2187 YSG candidates and 1949 RSG candidates from UCAC3 catalog. \citet{Neugent:2012aa} could however observe only 1528 YSG candidates and 865 RSG candidates in their 64 fields of the LMC, each of which has a $40$ arcmin field of view of multi-fiber, multi-object spectrometer \textit{Hydra}. These observed numbers are a fraction (70\% for YSG and 44\% for the RSGs) of the larger number of candidates in the UCAC3 catalog mentioned above. They discovered that 4\% of the YSG candidates and about 59\% of RSG candidates yielded nothing but sky upon examining the spectra.} %{\textbf{Note for the referee: The preceding deleted text has been moved to section \ref{subsec:neugents_criteria}, and is shown in orange.}} 
\deleted{ However, these cannot be spurious sources. About 93 \% of these RSG candidates have both proper motions measured from the UCAC3 and 2MASS photometry.  These candidates fall in the region of large J$-$K values (> 1.2 mag), i.e they are very cool stars (\Teff $<$ 3500 K). The authors concluded that many of these stars showed evidence of being highly dust-enshrouded and observed. Since the Hydra MOS spectroscopic pass band falls in the $I$-band, a typical M0 RSG with \Teff = 3800 K and $V_0 = 13.5$ and with the normal amount of LMC reddening $E(B-V) = 0.13$, can be expected to have $V \sim 14.0$ mag, $K\sim 9.6$ mag and $ I \sim 12.0$ mag. However if the star is shrouded in a thick circumstellar dust shell of a normal reddening law, resulting in $J-K =1.8$ mag, with $E(J-K) =0.8$ (and hence $A_V = 4.9$ mag instead of the normal $A_V =0.5$ mag), the star will be 3 mag fainter in the spectral passband $I$ (or at about $I = 15$ mag) than the normal stars for which good data is obtained with their exposures. These stars will therefore be at the faintest end of detectability. Excluding these highly reddened stars, 3\% of their RSG candidates yielded nothing but sky. } %{\textbf{Note for the referee: The preceding deleted text has been moved to section \ref{subsec:dust-enshrouded_RSG}, and is shown in orange.}} 
\deleted{The published Vizier catalog has 1447 YSG candidates and 521 RSG candidates. This is 95\% YSGs and 60\% RSG candidates initially observed by \citet{Neugent:2012aa} .}
%{\textbf{Note for the referee: The preceding deleted text has been moved to section \ref{subsec:neugents_criteria}, and is shown in orange.}}
%Their listing of the unique YSG and RSG {\it candidates} in the LMC are smaller than the subsample of the UCAC catalog because their 64 fields covered less sky than that of the %UCAC3 catalog

The spectroscopic observations \deleted{\added{of \citeauthor{Neugent:2012aa}} used Ca \textsc{ii}  triplet ($\lambda\lambda\lambda$ 8498, 8542, 8662)} to measure the radial velocities of the stars to \replaced{investigate the}{discriminate their} membership of the LMC \added{involved the use of Ca \textsc{ii}  triplet lines ($\lambda\lambda\lambda$ 8498, 8542, 8662) \citep{Neugent:2012aa}.} Taking account of the large radial velocity separation between the LMC and Milky Way stars, they found that 309 \added{of 1447} YSG candidates and 506 \added{of 865} RSG candidates \added{observed by them} have radial velocities higher than 200 \kmps \ and are probable LMC supergiants (labeled as category 1 stars \added{in the Vizier catalog}).  They found 8 candidates, all of which are YSGs, have 155 \kmps $ < v_{rad} <$ 200 \kmps . They labeled these stars as category 2 (possible but not probable) LMC supergiants. We use the category 1 stars listed in the Vizier catalog \deleted{(309 YSGs and 506 RSGs)} giving a total of 815 YSG+RSG candidates \added{in the LMC}.% for the entire luminosity and \lgTeff \ range.

\subsection{Blue, Yellow, and Red Supergiant candidate selection for comparison with our models}\label{subsec:our_criteria}

\deleted{Nevertheless, since} The blue/yellow to red ratio of supergiants in the LMC has been debated in the literature over a long time and is critical for comparing with model comparisons. We recapitulate below briefly the multiple filters by which the number of observed supergiants are identified and selected \added{for model comparison}.
As already mentioned, \citet{Neugent:2012aa} combine together B and A type stars with F and G type stars into their broadly defined ``YSG'' category. When we distinguish
stars with \lgTeff $ >$ 3.875 as BSGs, \added{those with} 3.875 $\ge $ \lgTeff $ >$ 3.68 as YSGs, \added{and those with} \lgTeff $ <$ 3.68 as RSGs, we get the candidates redistributed in to 163 BSG, 109 YSG, and 543 RSG candidates in these intervals (a total of 815 supergiants). The number of observed BSG, YSG and RSGs arising from various selection criteria relevant to our model calculations are summarized in our Table \ref{tab:supergiants}. Since \citet{Neugent:2012aa} are confident about completeness of the data only upto 4.0 $\ge $\lgTeff , we further report the restricted number of the BSG candidates in row 5 of Table \ref{tab:supergiants}\footnote{We note that their procedure of selection of supergiant candidates in the LMC was designed to make the YSG sample complete for stellar tracks of initial mass 12  \msun \ or higher}. Finally, since we are interested in comparing predictions of our model calculations with those of a specific set of mass ranges and overshoot factors, that display the Blue Loops, we report in last two rows of the table the observed number of BSGs, YSGs and RSGs with above temperature ranges in specific luminosity bands, which are discussed in section \ref{sec:BR_calc}.

%We use the following selection criteria based on the effective temperature to subdivide the observational data and model time-scales in to the blue, yellow, and red supergiant (B/Y/RSG) phases \citep[based on criterion used for YSG candidates in][]{Neugent:2012aa}. We use the same criteria to calculate the B/R and Y/R ratios using time-scales predicted by our models in each of these phases. After the terminal-age main sequence (TAMS) phase in the stellar evolution in our models, we consider a star to be in the BSG phase if 4.2 $\ge $\lgTeff $ >$ 3.875, in the YSG phase if 3.875 $\ge $ \lgTeff $ >$ 3.68, and in the RSG phase if 3.68 $\ge$ \lgTeff \  . However, we note here that the observational data from \citet{Neugent:2012aa} is restricted to \Teff \ $\lesssim$ 10,000 K (\lgTeff \ $\approx$ 4.0). Hence, we define a shorter BSG$^*$ phase for 4.0 $\ge $\lgTeff $ >$ 3.875. Using this modified BSG criterion and above Y/RSG criteria, we get a total of 112 BSG, 109 YSG, and 543 RSG candidates in the LMC from the \cite{Neugent:2012aa} data. We note that there are additional 26 BSG candidates in the observational data that have \lgL \ $<$ 4.0, which could be blue dwarfs or main-sequence stars. We omitted from our counts, as they are too dim in comparison to our models. 

\subsection{Dust Enshrouded RSGs}\label{subsec:dust-enshrouded_RSG}

\added{As discussed in section 3.1, \citet{Neugent:2012aa} discovered after examining the spectra that 69 YSGs and 343 RSGs out of the observed candidates shortlisted from UCAC3 catalog yielded nothing but sky background, i.e. there were no detections of these RSGs. While a few YSGs can be missed due to low transmission in some fibers, the number of missing RSGs is too high.}\deleted{However,} \added{All of} these \added{UCAC3} RSGs cannot be spurious sources. About 93\% of these \added{343} RSG candidates have both proper motions measured from the UCAC3 and 2MASS photometry. These candidates fall in the region of large J$-$K values (> 1.2 mag), i.e they are very cool stars (\Teff $<$ 3500 K). The authors \replaced{concluded}{conjectured} that many of these stars showed evidence of being highly dust-enshrouded\deleted{ and observed}. Since the Hydra MOS spectroscopic pass band falls in the $I$-band, a typical M0 RSG with \Teff = 3800 K and $V_0 = 13.5$ and with the normal amount of LMC reddening $E(B-V) = 0.13$, can be expected to have $V \sim 14.0$ mag, $K\sim 9.6$ mag and $ I \sim 12.0$ mag \added{which would have been easily detected with the given exposure}. However if the star is shrouded in a thick circumstellar dust shell of a normal reddening law, resulting in $J-K =1.8$ mag, with $E(J-K) =0.8$ (and hence $A_V = 4.9$ mag instead of the normal $A_V =0.5$ mag), the star will be 3 mag fainter in the spectral passband $I$ (or at about $I = 15$ mag) than the normal stars for which good data is obtained with their exposures. These stars will therefore be at the faintest end of detectability. \added{Thus corresponding to each of the 543 detected RSGs, there may be roughly
60 \% more hidden RSGs in the observed LMC fields that could be shrouded and obscured in dust.} \deleted{After excluding these highly reddened stars, only 3\% of the RSG candidates observed by \citeauthor{Neugent:2012aa} yielded nothing but sky.}

%%%% Table %%%%%%
\begin{longrotatetable}
\begin{deluxetable*}{cccccccccccccccc}
\tablecaption{Model Lifetimes for RSG/YSG/BSG phases \label{tab:br_ratios}}
\tablewidth{0pt}
\tabletypesize{\scriptsize}
\tablehead{
\colhead{overshoot} & \colhead{$\rm \tau_{B_1}$} & 
\colhead{$\rm \tau_{Y_1}$} & \colhead{$\rm \tau_{R_1}$} &
\colhead{$\rm \tau_{Y_2}$} & \colhead{$\rm \tau_{B_2}$} & 
\colhead{$\rm \tau_{Y_3}$} & \colhead{$\rm \tau_{R_2}$} & 
\multicolumn{4}{c}{Total Model Timescale  in \powten{3} years} &
\multirow{2}{*}{$\rm \tau_{BSG^{*}} / \tau_{RSG}$} & 
\multirow{2}{*}{$\rm \tau_{BSG} / \rm \tau_{RSG}$} & 
\multirow{2}{*}{$\rm \tau_{YSG} / \rm \tau_{RSG}$} &
\colhead{$\rm log_{10} (L,T_{eff})$} \\ [-1ex] 
\colhead{parameter, f} & \colhead{(BSG 1)} & \colhead{(YSG 1)} & 
\colhead{(RSG 1)} & \colhead{(YSG 2)} & \colhead{(BSG 2)} &
\colhead{(YSG 3)} & \colhead{(RSG 2)} & 
\colhead{BSG$^{*}$} & \colhead{BSG} & \colhead{YSG} & \colhead{RSG} &
\colhead{} & \colhead{} & \colhead{} & \colhead{at T.P.}
}
%\colnumbers
\startdata
\multicolumn{13}{c}{\textbf{12 \msun}} \\
\hline
0.019
& 11.64 & \phn 2.43 & \phn 672.51 & 5.98 & 652.96 & \phn 9.31 & \phn 69.63 & \phn 41.92 & 664.60 & \phn 17.71 & \phn 742.15 & 0.06\phn & 0.90\phn & 0.02\phn & (4.60, 4.13) \\
0.020
& 11.26 & \phn 2.34 & \phn 703.07 & 6.23 & 594.41 & 12.91 & 70.65 & \phn 53.99 & 605.67 & \phn 21.47 & \phn 773.72 & 0.07\phn & 0.78\phn & 0.03\phn & (4.60, 4.12) \\
0.021
& 10.82 & \phn 2.29 & \phn 813.34 & 7.24 & 453.03 & 17.68 & 73.24 & 102.86 & 463.85 & \phn 27.21 & \phn 886.58 & 0.12\phn & 0.52\phn & 0.03\phn & (4.61, 4.05) \\
0.022
& 10.42 & \phn 2.25 & \phn 865.63 & 8.09 & 367.99 & 26.56 & 77.42 & 265.87 & 378.42 & \phn 36.90 & \phn 943.06 & 0.28\phn & 0.40\phn & 0.04\phn & (4.61, 4.01) \\
0.023
& 10.07 & \phn 2.22 & 1324.79 & -- & -- & -- & -- & \phn \phn 7.65 & \phn 10.07 & \phn \phn 2.22 & 1324.79 & 0.006 & 0.008 & 0.002 & N/A \\
0.024
& \phn 9.87 & \phn 2.15 & 1304.95 & -- & -- & -- & -- & \phn \phn 7.53 & \phn \phn 9.87 & \phn \phn 2.15 & 1304.95 & 0.006 & 0.008 & 0.002 & N/A \\
0.025
& \phn 9.52 & \phn 2.10 & 1283.30 & -- & -- & -- & -- & \phn \phn 7.24 & \phn \phn 9.52 & \phn \phn 2.10 & 1283.30 & 0.006 & 0.007 & 0.002 & N/A \\
0.030
& \phn 8.05 & \phn 1.90 & 1186.52 & -- & -- & -- & -- & \phn \phn 6.09 & \phn \phn 8.05 & \phn \phn 1.90 & 1186.52 & 0.005 & 0.007 & 0.002 & N/A \\
0.031
& \phn 7.87 & \phn 1.87 & 1171.15 & -- & -- & -- & -- & \phn \phn 5.95 & \phn \phn 7.87 & \phn \phn 1.87 & 1171.15 & 0.005 & 0.007 & 0.002 & N/A \\
0.050
& \phn 4.74 & \phn 1.27 & \phn 948.84 & -- & -- & -- & -- & \phn \phn 3.54 & \phn \phn 4.74 & \phn \phn 1.27 & \phn 948.84 & 0.004 & 0.005 & 0.001 & N/A \\
\hline
\multicolumn{13}{c}{\textbf{13 \msun}} \\
\hline
0.015
& 11.23 & \phn 2.44 & 1367.13 & -- & -- & -- & -- & \phn \phn 8.62 & \phn 11.23 & \phn \phn 2.44 & 1367.13 & 0.006 & 0.008 & 0.002 & N/A \\
0.017
& 10.37 & \phn 2.20 & 1321.51 & -- & -- & -- & -- & \phn \phn 8.01& \phn 10.37 & \phn \phn 2.20 & 1321.51 & 0.006 & 0.008 & 0.002 & N/A \\
0.018
& 10.03 & \phn 2.14 & \phn 642.59 & 4.55 & 536.94 & 20.58 & 85.95 & \phn 66.87 & 546.97 & \phn 27.26 & \phn 728.54 & 0.09\phn & 0.75\phn & 0.04\phn & (4.67, 4.08) \\
0.020
& \phn 9.40 & \phn 1.98 & \phn 351.03 & 3.96 & 816.07 & \phn 5.64 & 53.22 & \phn 22.91 & 825.47 & \phn 11.58 & \phn 404.25 & 0.06\phn & 2.04\phn & 0.03\phn & (4.69, 4.14) \\
0.025
& \phn 7.84 & \phn 1.75 & \phn 431.04 & 4.18 & 648.10 & \phn 8.18 & 46.16 & \phn 31.80 & 655.94 & \phn 14.11 & \phn 477.21 & 0.07\phn & 1.37\phn & 0.03\phn & (4.72, 4.13) \\
0.030 
& \phn 6.85 & \phn 1.57 & \phn 542.97 & 6.03 & 429.05 & 20.19 & 58.16 & \phn 84.69 & 435.90 & \phn 27.79 & \phn 601.13 & 0.14\phn & 0.73\phn & 0.05\phn & (4.75, 4.08) \\
0.031
& \phn 6.62 & \phn 1.56 & \phn 587.07 & 6.94 & 352.29 & 27.54 & 69.52 & \phn 164.71 & 358.91 & \phn 36.04 & \phn 656.59 & 0.25\phn & 0.55\phn & 0.05\phn & (4.75, 4.03) \\
0.032
& \phn 6.39 & \phn 1.52 & 1030.23 & -- & -- & -- & -- & \phn \phn 4.82 &  \phn \phn 6.39 & \phn \phn 1.52 & 1030.23 & 0.005 & 0.006 & 0.001 & N/A \\
0.033
& \phn 6.17 & \phn 1.49 & 1017.68 & -- & -- & -- & -- & \phn \phn 4.67 &\phn \phn  6.17 & \phn \phn 1.49 & 1017.68 & 0.005 & 0.006 & 0.001 & N/A \\
0.035
& \phn 5.83 & \phn 1.42 & \phn 993.70 & -- & -- & -- & -- & \phn \phn 4.38 & \phn \phn 5.83 & \phn \phn 1.42 & \phn 993.70 & 0.004 & 0.006 & 0.001 & N/A \\
0.050
& \phn 4.10 & \phn 1.08 & \phn 859.71 & -- & -- & -- & -- & \phn \phn 3.07 & \phn \phn 4.10 & \phn \phn 1.08 & \phn 859.71 & 0.004 & 0.005 & 0.001 & N/A \\
\hline
\multicolumn{13}{c}{\textbf{14 \msun}} \\
\hline
0.020
& \phn 9.20 & \phn 2.54 & 1115.81 & -- & -- & -- & -- & \phn \phn 6.93 & \phn \phn 9.20 & \phn \phn 2.54 & 1115.81 & 0.006 & 0.008 & 0.002 & N/A \\
0.025
& \phn 7.62 & \phn 1.71 & 1046.03 & -- & -- & -- & -- & \phn \phn 5.85 & \phn \phn 7.62 & \phn \phn 1.71 & 1046.03 & 0.006 & 0.007 & 0.002 & N/A \\
0.027
& \phn 6.89 & \phn 1.54 & 1010.78 & -- & -- & -- & -- & \phn \phn 5.29 & \phn \phn 6.89 & \phn \phn 1.54 & 1010.78 & 0.005 & 0.007 & 0.002 & N/A \\
0.028 
& \phn 6.53 & \phn 1.47 & \phn 296.46 & 3.32 & 622.49 & 13.90 & 48.31 & \phn 41.16 & 629.02 & \phn 18.70 & \phn 344.78 & 0.12\phn & 1.82\phn & 0.05\phn & (4.82, 4.12) \\
0.030
& \phn 6.17 & \phn 1.43 & \phn 333.93 & 3.83 & 539.87 & 22.31 & 62.99 & \phn 66.98 & 546.04 & \phn 27.57 & \phn 396.92 & 0.17\phn & 1.38\phn & 0.07\phn & (4.83, 4.07) \\
0.031
& \phn 6.01 & \phn 1.39 & \phn 303.10 & 3.38 & 580.39 & 14.76 & 46.43 & \phn 42.84 & 586.40 & \phn 19.53 & \phn 349.53 & 0.12\phn & 1.68\phn & 0.06\phn & (4.83, 4.11) \\
0.033
& \phn 5.58 & \phn 1.31 & \phn 368.22 & 4.20 & 449.65 & 28.60 & 73.75 & \phn 87.53 & 455.24 & \phn 34.12 & \phn 441.98 & 0.20\phn & 1.03\phn & 0.08\phn & (4.84, 4.06) \\
0.035
& \phn 5.49 & \phn 1.30 & \phn 332.30 & 3.95 & 500.32 & 19.50 & 49.60 & \phn 56.68 & 505.81 & \phn 24.74 & \phn 381.90 & 0.15\phn & 1.32\phn & 0.06\phn & (4.85, 4.12) \\
0.037 
& \phn 5.37 & \phn 1.26 & \phn 331.26 & 4.21 & 472.42 & 23.08 & 56.63 & \phn 67.84 & 477.79 & \phn 28.56 & \phn 387.89 & 0.17\phn & 1.23\phn & 0.07\phn & (4.86, 4.10) \\
0.040
& \phn 4.61 & \phn 1.16 & \phn 862.05 & -- & -- & -- & -- & \phn \phn 3.49 & \phn \phn 4.61 & \phn \phn 1.16 & \phn 862.05 &  0.004 & 0.005 & 0.001 & N/A \\
0.050
& \phn 3.66 & \phn 0.98 & \phn 798.09 & -- & -- & -- & -- & \phn \phn 2.76 & \phn \phn 3.66 & \phn \phn 0.98 & \phn 798.09 & 0.003 & 0.005 & 0.001 & N/A \\
\hline
continued ... \\
\\
\multicolumn{13}{c}{\textbf{15 \msun}} \\
\hline
0.035
& \phn 5.84 & \phn 1.41 & \phn 844.18 & -- & -- & -- & -- & \phn \phn 4.42 & \phn \phn 5.84 & \phn \phn 1.41 & \phn 844.18 & 0.005 & 0.007 & 0.002 & N/A \\
0.037
& \phn 5.50 & \phn 1.29 & \phn 830.71 & -- & -- & -- & -- & \phn \phn 4.16 & \phn \phn 5.50 & \phn \phn 1.29 & \phn 830.71 & 0.005 & 0.007 & 0.002 & N/A \\
0.040
& \phn 4.47 & \phn 1.16 & \phn 277.14 & 3.63 & 411.62 & 31.92 & 78.78 & \phn 85.80 & 416.09 & \phn 36.71 & \phn 355.91 & 0.24\phn & 1.17\phn & 0.10\phn & (4.94, 4.08) \\
0.045 
& \phn 4.14 & \phn 1.16 & \phn 260.02 & 4.14 & 398.43 & 30.30 & 71.21 & \phn 80.93 & 402.57 & \phn 35.60 & \phn 331.23 & 0.24\phn & 1.21\phn & 0.10\phn & (4.96, 4.10) \\
0.047 
& \phn 3.80 & \phn 0.98 & \phn 763.04 & -- & -- & -- & -- & \phn \phn 2.89 & \phn \phn 3.80 & \phn \phn 0.98 & \phn 763.04 & 0.004 & 0.005 & 0.001 & N/A \\
0.050
& \phn 3.59 & \phn 0.98 & \phn 753.16 & -- & -- & -- & -- & \phn \phn 2.67 & \phn \phn 3.59 & \phn \phn 0.98 & \phn 753.16 & 0.004 & 0.005 & 0.001 & N/A \\
\hline
\multicolumn{13}{c}{\textbf{18 \msun}} \\
\hline
0.040
& \phn 6.45 & \phn 2.38 & \phn 677.70 & -- & -- & -- & -- & \phn \phn 4.58 & \phn \phn 6.45 & \phn \phn 2.38 & \phn 677.70 & 0.007 & 0.01\phn & 0.004 & N/A \\
0.050
& \phn 3.28 & \phn 1.08 & \phn 613.28 & -- & -- & -- & -- & \phn \phn 2.38 & \phn \phn 3.28 & \phn \phn 1.08 & \phn 613.28 & 0.004 & 0.005  & 0.002 & N/A \\
\hline
\multicolumn{13}{c}{\textbf{20 \msun}} \\
\hline
0.035
& \phn 6.47 & \phn 3.21 & \phn 641.57 & -- & -- & -- & -- & \phn \phn 4.42 & \phn \phn 6.47 & \phn \phn 3.21 & \phn 641.57 & 0.007 & 0.01\phn & 0.005 & N/A \\
0.045
& \phn 4.48 & \phn 2.06 & \phn 593.42 & -- & -- & -- & -- & \phn \phn 3.30 & \phn \phn 4.48 & \phn \phn 2.06 & \phn 593.42 & 0.006 & 0.008 & 0.003 & N/A \\
0.050
& \phn 2.89 & \phn 1.09 & \phn 555.95 & -- & -- & -- & -- & \phn \phn 2.07 & \phn \phn 2.89 & \phn \phn 1.09 & \phn 555.95 & 0.004 & 0.005 & 0.002 & N/A \\
\hline
\multicolumn{13}{c}{\textbf{24 \msun}} \\
\hline
0.035
& 58.78 & 156.82 \phn & \phn 356.83 & -- & -- & -- & -- & 6.43 & 58.78 & 156.82 & \phn 356.83 & 0.02\phn & 0.16\phn & 0.44\phn & N/A
\enddata
\tablecomments{The time ($\tau$ in years) spent in each of the blue, yellow, and red supergiant (B/Y/RSG) phases are listed here for various M$_{ZAMS} $ models with Z = 0.006 described in this paper. All models have $f_0$ held at 0.005. The phases are divided on the basis of effective temperature ($\rm T_{eff}$) as follows - after the TAMS phase in the evolution of the star if 4.2 $\ge $\lgTeff $ >$ 3.875 then the star is considered to be in the BSG phase, if 3.875 $\ge $ \lgTeff $ >$ 3.68 then in the YSG phase, and 3.68 $\ge$ \lgTeff \ then in the RSG phase. The models that exhibit blue loops (f $<$ 0.032) go through each of these phases several times, during red-ward and blue-ward tracks. Each of these times are listed separately, when applicable. The total time spent in B/Y/RSG phases are also calculated and listed. Column BSG$^{*}$ shows time spent in BSG phase when 4.0 $\ge $\lgTeff $ >$ 3.875, to compare with the available observational data for BSG candidates. \added{The luminosity and effective temperature at the turning point (T.P.) in the blue loop are listed in the last column.} All of these models, with or without the blue loop, terminate in the RSG phase at CC.}
\end{deluxetable*}
\end{longrotatetable}

%%%%%%%%% TABLE %%%%%%%%%%%
%\begin{deluxetable}{ccccc}
%\tablecaption{Model timescales for selected models in $\rm 10^{3}$ years for 4.55 $\le $ \lgL $ \le $ 5.0 \label{tab:br_select}}
%%\tablewidth{0pt}
%%\tabletypesize{\scriptsize}
%\tablehead{
%\colhead{Star Mass} & \colhead{overshoot} & 
%\multicolumn{3}{c}{Total Model Timescale} 
%\\ [-2ex] 
%\colhead{$\rm M_{\odot}$} & \colhead{parameter, f} &
%\colhead{$\tau_{BSG}$} & \colhead{$\tau_{YSG}$} & \colhead{$\tau_{RSG}$}
%}
%%\colnumbers
%\startdata
%12 & 0.022 & 128.42 & 34.65 & 468.31 \\
%13 & 0.031 & \phn 85.17 & 35.57 & 655.33 \\
%14 & 0.037 & \phn 12.63 & 28.56 & 379.61 \\
%15 & 0.045 & \phn 18.56 & 35.60 & 264.45 \\
%\enddata
%\tablecomments{The model predicted supergiant time-scales ($\tau$) for 4.55 $\le $ \lgL $ \le $ 5.0, and 4.0 $\ge $\lgTeff $ >$ 3.875 for BSG, 3.875 $\ge $ \lgTeff $ >$ 3.7 for YSG, and 3.68 $\ge$ \lgTeff \ for RSG phases. These time-scales are used to calculate the B/R \& Y/R ratios using equation \ref{eqn:BtoR_mod}. The models listed here undergo a blue loop. Thus, the listed time-scales include several red-ward and blue-ward transitions in each of the B/Y/RSG phases.}
%\end{deluxetable}
%%%%%%%%%%%%%%%

%\subsection{}

%%%%%% Figure %%%%%
%\begin{figure*}[htb!]
%\includegraphics[width=0.98\textwidth]{Fig_all_masses.pdf}
%\caption{all masses
%\label{fig:all}}
%\end{figure*}

\begin{figure*}[htb!]
\includegraphics[width=0.98\textwidth]{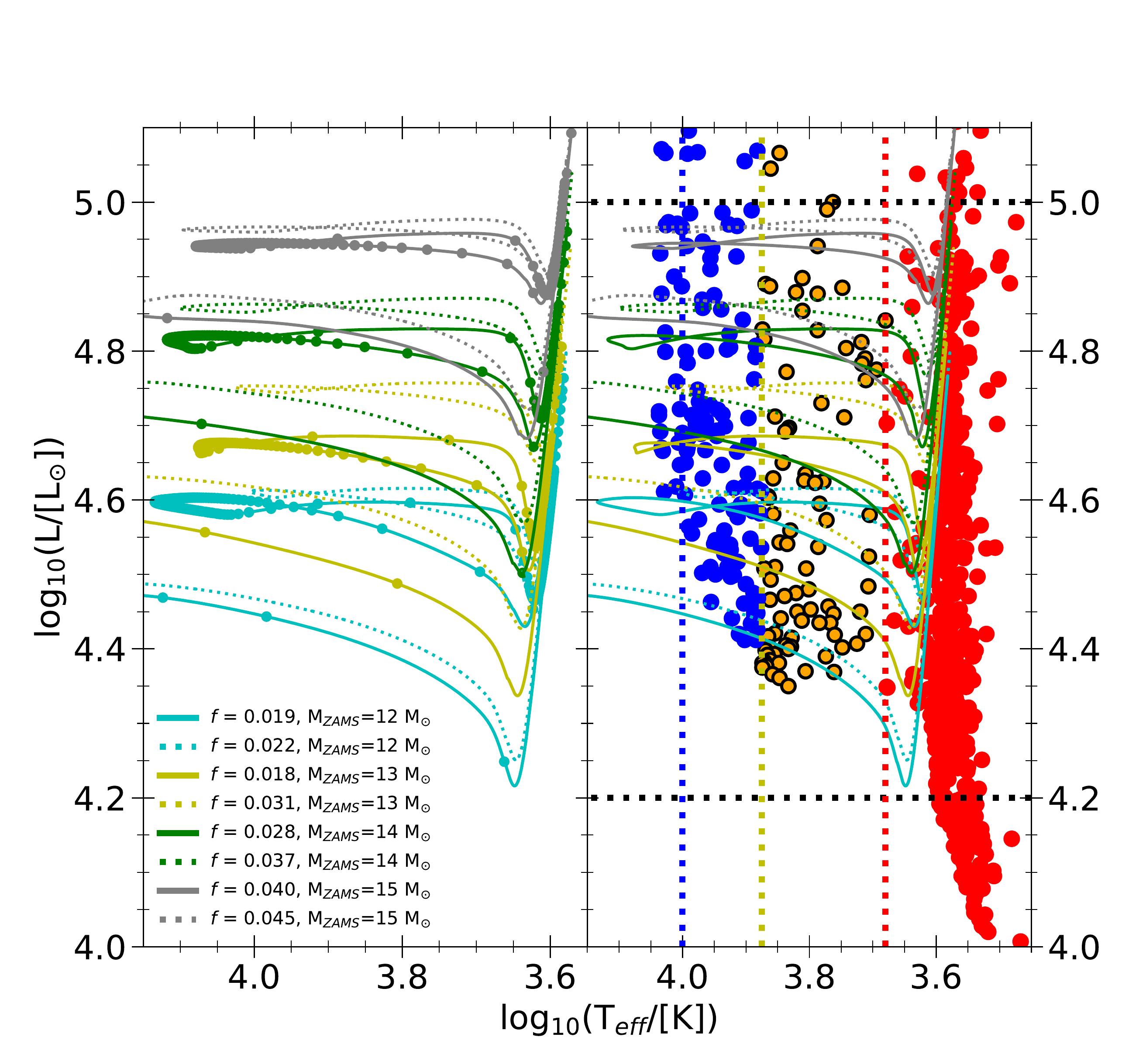}
\caption{The HR diagram for models in the range of 12 to 15 \msun \ with minimum (solid line) and maximum (dotted line) $f$ values for each mass that undergo a blue loop. The panel on the left shows evolutionary tracks marked with circles at every 5000 years period. The same tracks are shown in the right panel with the observational data from \citet{Neugent:2012aa} overlaid on them. The blue, yellow, and red supergiant (B/Y/RSG) candidates divided based on their effective temperature as described in section \ref{sec:obs_HR}. The vertical dotted lines \added{in the right panel} show the temperature limits \added{of the B/Y/RSG}. The horizontal black dotted lines show the selection \replaced{upper}{of} luminosity limits for comparison of the observational data with our models.
\label{fig:hrd}}
\end{figure*}

%\begin{figure*}[htb!]
%\includegraphics[width=0.98\textwidth]{Fig1.pdf}
%\caption{selected models
%\label{fig:12msun}}
%\end{figure*}
%
%\begin{figure*}[htb!]
%\includegraphics[width=0.98\textwidth]{Fig2.pdf}
%\caption{selected models
%\label{fig:14msun}}
%\end{figure*}
%
%\begin{figure*}[htb!]
%\includegraphics[width=0.98\textwidth]{Fig3.pdf}
%\caption{selected models
%\label{fig:15msun}}
%\end{figure*}
%
%\begin{figure*}[htb!]
%\includegraphics[width=0.98\textwidth]{Fig4.pdf}
%\caption{selected models
%\label{fig:18msun}}
%\end{figure*}
%
%\begin{figure*}[htb!]
%\includegraphics[width=0.97\textwidth]{Fig5.pdf}
%\caption{selected models
%\label{fig:20msun}}
%\end{figure*}
%
%\begin{figure*}[htb!]
%\includegraphics[width=0.97\textwidth]{Fig6.pdf}
%\caption{None of the models have converged for 24 \msun .
%\label{fig:24msun}}
%\end{figure*}
%
\begin{figure}[htb!]
\includegraphics[width=0.45\textwidth]{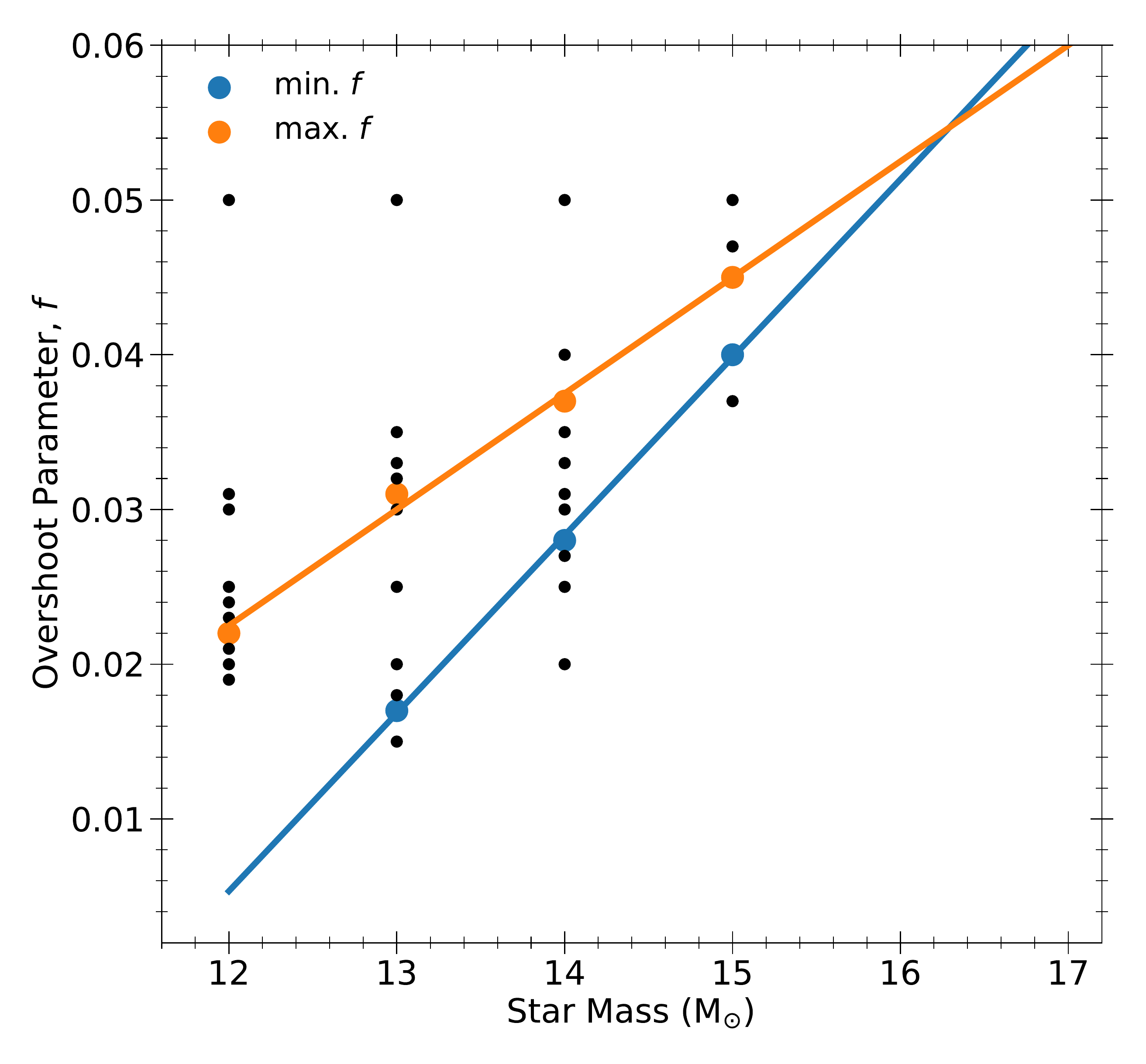}
\caption{The grid of \added{overshoot parameter} $f$ values for each mass \added{used in computations in this work} is shown with circles. The orange and blue circles show the maximum and minimum $f$ values that undergo a blue loop, respectively. The two fitted lines show that these $f$ values \replaced{are linear in}{increase linearly with} mass. The models with masses larger than about 16 \msun , where the two lines intersect do not undergo a blue loop for any choice of the $f$ value\deleted{, which is also evident from the models that we ran}.
\label{fig:ov_range}
}
\end{figure}
%
%\begin{figure*}[htb!]
%\includegraphics[width=0.97\textwidth]{Fig9.pdf}
%\caption{The values at TAMS stage of log L and \Teff \ plotted as a function overshoot parameter $f$ for each mass.
%\label{fig:tamsLT}}
%\end{figure*}

%\begin{figure*}[htb!]
%\gridline{\fig{Fig10.pdf}{0.45\textwidth}{a}
%				\fig{Fig11.pdf}{0.45\textwidth}{b}}
%\caption{The values at TAMS stage of log L plotted as a function overshoot parameter $f$ for each mass.
%\label{fig:tamsL_3d}}
%\end{figure*}

%\begin{figure*}[htb!]
%\includegraphics[width=0.97\textwidth]{Fig11.pdf}
%\caption{The values at TAMS stage of log L plotted as a function overshoot parameter $f$ for each mass.
%\label{fig:tamsL}}
%\end{figure*}

%\begin{figure*}[htb!]
%\includegraphics[width=0.97\textwidth]{Fig12.pdf}
%\caption{B/R time scales plotted at each overshoot parameter $f$ for each mass.
%\label{fig:BR}}
%\end{figure*}

%\begin{figure*}[htb!]
%\includegraphics[width=0.97\textwidth]{Fig13.pdf}
%\caption{Y/R time scales plotted at each overshoot parameter $f$ for each mass.
%\label{fig:YR}}
%\end{figure*}
%%%%%%%%%%%%%%%

\begin{figure}[htb!]
\includegraphics[width=0.49\textwidth]{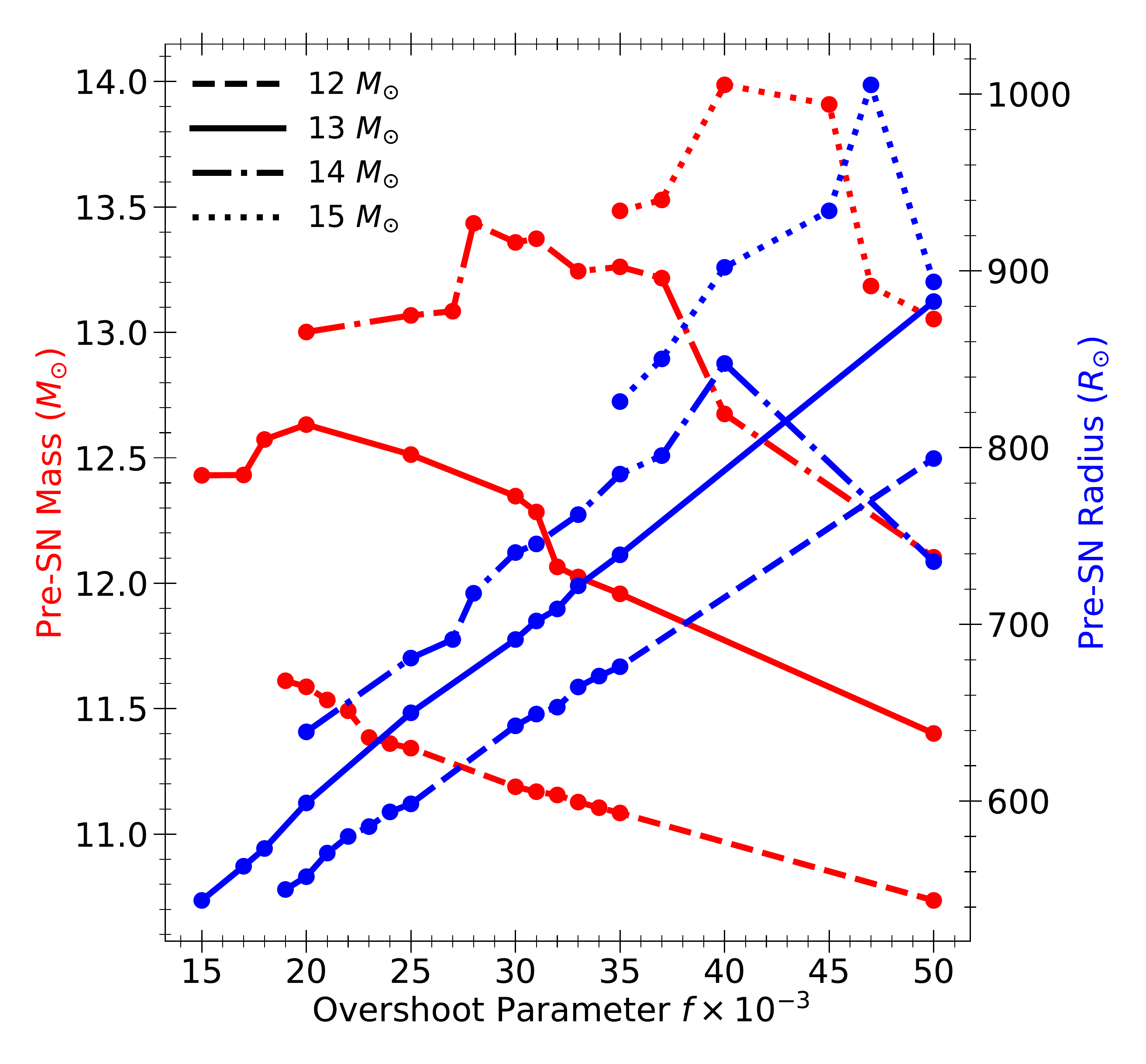}
\caption{The final pre-supernova masses and radii as a function of overshoot parameter plotted for different zero-age main-sequence masses in the range of 12--15 \msun . 
\label{fig:mass_variation}
}
\end{figure}

\section{Method of Simulation of Massive Star Evolution} \label{sec:methods}

We use MESA version r-10398 to explore models of progenitor stars of type IIP supernovae, such as SN 2013ej. In our Paper I, we explored effect of overshoot parameter $f$ on stellar properties and evolution for a \added{single, non-rotating} model star with initial mass of 13 \msun \ at ZAMS, with initial metallicity Z = 0.006. In this paper we expand the parameter space \added{of the same type of stars}. In addition to the 13 \msun \ star (see Paper I for details), we use an expanded grid of stellar masses and overshoot parameter, $f$ (see Fig \ref{fig:ov_range} and Table \ref{tab:br_ratios} for the grid points of mass and exponential overshoot factor $f$ explored). We use the same metallicity, mass and temporal resolution controls, mass loss rate ($\eta_{Dutch}$=0.5), nuclear reaction network of 79 isotopes, convection with Ledoux criterion ($\alpha_{MLT}$ = 2), and semiconvection ($\alpha_{SC}$ = 0.1) as in the case of the 13 \msun \ star in Paper I. We vary the overshoot parameter for exponential decay $f$, keeping the value of other parameter $f_0$ constant at 0.005. We have listed all of the models between a mass range of 12--24 \msun \ and a range of values of overshoot parameter $f$ for each mass in Table \ref{tab:br_ratios}. Only the models that have converged from the pre-main-sequence to either the core-collapse stage or within a day to core-collapse stage (which does not affect the overall RSG lifetimes) are listed in this Table. Some of the models that failed to converge soon after core-He-depletion stage due to various reasons were not useful for calculating the red supergiant time-scales to determine the B/R ratio and are thus not reported in this Table.

\section{Results}\label{sec:results}

\subsection{Model time-scales in the supergiant phases and B/R ratio calculation} \label{sec:BR_calc}

In Fig. \ref{fig:hrd}, we have plotted the model evolutionary tracks in the mass range of 12 to 15 \msun \ with maximum and minimum model $f$-values that undergo a blue loop for each mass (see Table \ref{tab:br_ratios}). The observational data is overlaid on the tracks in the right panel of Fig. \ref{fig:hrd}. The two horizontal (black-dashed) lines show the luminosity range used to restrict the observational data for comparison with the model calculations of B/R and Y/R ratios. We choose this range of 4.2 $<$ \lgL $<$ 5.0 for limiting the observational data, as this is the range covered by the model evolutionary tracks that is most appropriate for the comparison. %The vertical RSG tracks above this luminosity for the 15 \msun \ models do not contribute significantly to the total RSG time-scales, and hence, we can safely save that we have not rejected a significant number of RSG with initial mass $<$ 15 \msun \ from the observational data by setting this upper limit. 
With this luminosity limit we have 97 BSG, 87 YSG and 430 RSG candidates as noted in the last row of Table \ref{tab:supergiants}. These values are used to compute the observed B/R and Y/R ratios listed in Table \ref{tab:br_comparison}.

In Table \ref{tab:br_ratios}, we list the time-scales calculated for our models in the B/Y/RSG phases as defined in section \ref{sec:obs_HR}. The models that undergo a blue loop go through each of these phases several times, red-ward and blue-ward. $\rm \tau_{B_1}$, $\rm \tau_{Y_1}$, $\rm \tau_{R_1}$ time-scales represent the first time after TAMS phase when the star is in the B/Y/RSG phases, respectively, on the red-ward track in the HRD. For the models that do not undergo a blue loop, these are also the total time-scales spent in each of these phases. For the blue loop models, several such time-scales are listed in the table for each consecutive transition. $\rm \tau_{Y_2}$ represents the time-scale in YSG phase when the star is on the blue-ward track on the blue loop. $\rm \tau_{B_2}$ represents the time-scale for the BSG phase when the star is in the blue region (\lgTeff \ $>$ 3.875). This includes both the red-ward and blue-ward tracks including the loop. $\rm \tau_{Y_3}$ represents the time-scale when the star is in the YSG phase on the red-ward track in the blue loop. The star then enters the RSG phase and remains in this \replaced{regime}{spectral class} until core-collapse stage. This time-scale is represented by $\rm \tau_{R_2}$. The total time-scales in each of the B, Y, \& RSG phases are then listed as the sum of the above individual time-scales in columns BSG, YSG, and RSG, respectively. The ratios of the total time-scales in the BSG to the RSG and the YSG to the RSG phases are also listed in Table \ref{tab:br_ratios} for comparison in the columns $\rm \tau_{BSG}/\tau_{RSG}$ \& $\rm \tau_{YSG}/\tau_{RSG}$, respectively. Since the observational data from \citet{Neugent:2012aa} is complete only up to \Teff \ $\approx$ 10,000 K, we also calculate BSG time-scales in the limit of 4.0 $\ge $\lgTeff $ >$ 3.875 for each of our models for comparison with the observations. The total BSG time-scales in this restricted temperature range are marked as BSG$^*$ in Table \ref{tab:br_ratios}. The corresponding ratio of individual blue to red time-scales is also listed in the column $\rm \tau_{BSG^*}/\tau_{RSG}$.

We note that in the \citet{Neugent:2012aa} data, the LMC stars are predominantly field stars. Since the field stars do not co-evolve like those in the clusters, we can assume a stellar population in a steady state with a constant star formation rate. We can thus evaluate the B/R ratio in terms of the relative duration of the two supergiant phases. For constant star formation rate, the B/R ratio \citep[$N_B/N_R$,]{Ekstrom:2013aa} can be modified for discrete mass values as :
\begin{equation} \label{eqn:BtoR}
\frac{N_B}{N_R} = \frac{\int\limits_{m} \tau_B(m) \ \phi(m) \ dm}{ \int\limits_{m} \tau_R(m) \ \phi(m) \ dm}
\end{equation}
where $ \phi(m) = (dN/dm) \propto m^{-\alpha}$ is the initial mass function with slope $\alpha = -$2.35 (Salpeter, 1995). However, we find from our model predictions that the supergiant time-scales, and hence, the B/R ratio are also dependent on the convective overshoot parameter $f$ for a given mass (see Table \ref{tab:br_ratios} for time-scales). Thus we introduce the $f$-dependence of the ratio in equation \ref{eqn:BtoR} as
\begin{equation} \label{eqn:BtoR_mod}
\frac{N_B}{N_R} = \frac{\int\limits_{m} \phi(m) \ dm \int\limits_{f} \tau_B(m,f) \ P(f)|_{m} \ d f}{ \int\limits_{m} \phi(m) \ dm \int\limits_{f} \tau_R(m,f) \ P(f)|_{m} \ df}
\end{equation}
where $ P(f)|_m$ is a probability function for finding stars with a particular $f$ for a given mass. The nature of convective overshoot and its mass dependence is not known, and hence, $P(f)$ cannot be well determined. As trials, we use two different forms of the probability functions to calculate the B/R \& Y/R ratios -- (1) $P(f) = constant$ (i.e. all $f$ values are equally probable) and (2) a \textit{normalized} Gaussian of the form 
\begin{equation} \label{eqn:P_Gaussian}
P(f) = \frac{1}{\sqrt{ 2 \pi \Delta^2 }} e^{ - ((f - f_c)/ 2 \Delta)^2 }
\end{equation}
where,  $f_c$ is the ``central'' value and $\Delta$ is the ``width'' of the Gaussian. 

We see from Figure \ref{fig:ov_range} that the models that exhibits a blue loop are bound between two extreme values of the overshoot parameter, $f$. Furthermore, these two extreme values appear to vary linearly with mass (orange and blue lines in the figure). We see that these two lines intersect at around 16.3 \msun , above which the models would not undergo a blue loop for any $f$ value in our selected range. As pointed earlier, the models that undergo a blue loop spend a significant amount of time in the BSG phase, compared to the models that do not undergo a blue loop. Hence, the number of BSG stars seen in the observations would be dominated by the models that undergo a blue loop. We choose the central value $f_c(m)$ for the Gaussian probability function $P(f)|_m$ at the maximum $f$ value\footnote{We note that the maximum $f$ value for blue looping models gives the maximum $\rm \tau_{BSG^{*}}$ time-scale and $\rm \tau_{BSG^{*}}$/$\rm \tau_{RSG}$ ratio for each mass, with an exception of 14 \msun . However, for 14 \msun \ these values are close to the maximum values.} for each mass that exhibits a blue loop (orange circles in Figure \ref{fig:ov_range}), which follows the (orange) line such that 
$$f_c(m) = 0.0075 m - 0.0675$$
We choose the $\Delta$ (``width'') for the Gaussian to be equal to half of the difference between the minimum and maximum $f$ values (orange and blue circles in Figure \ref{fig:ov_range}) for each mass. We evaluate the \textit{normalized} Gaussian probability function $P(f)$ over a range of $f$ values between 0.01 and 0.05. However, we only use the time-scales from the models that undergo a blue loop for the ratio calculation, for the reason mentioned above. 

To calculate the B/R ratio, we use the trapezoidal method to evaluate the integral in equation \ref{eqn:BtoR_mod} for discrete values of $m$ and $f$. We get B/R = 0.15 for both the constant $P(f)$ and the Gaussian $P(f)$.  We also calculate Y/R ratio using same methodology to yield Y/R = 0.044 and 0.058, respectively, for constant and Gaussian $P(f)$. We also compute the ratio of blue+yellow supergiants to red supergiants (B+Y)/R = 0.19 and 0.21, respectively, for constant and Gaussian $P(f)$. These values are listed in Table \ref{tab:br_comparison}. We infer from these results that the choice for the form of the probability function $P(f)$ has a minimal effect on the predicted B/R and Y/R ratios.

\added{We reiterate that the parameter f which
most significantly controls the convective overshoot and its effects on stellar mixing and evolution leading to the blue loop behavior in the HRD has a certain range that
appears to narrow with increasing (initial) mass $M$. The overshoot factor corresponding to blue loop behavior itself increases with $M$. Taken together with the influence of the blue loop on the population distribution of LMC supergiant stars in the HRD for the 12 to $\sim$16 \msun \ range this suggests that for more massive stars convective overshooting may be stronger, at least for a certain range of initial masses. We note that \citet{Castro:2014aa} using the spectroscopic HRD\footnote{Spectroscopic HRDs are independent of distance and extinction measurements but are based on spectroscopically derived effective temperatures and gravities calculated from stellar atmosphere modelling and spectra of the stars.} of nearly 600 massive stars in the
{\it Milky-way galaxy} suggest that model calculations require a mass dependent overshooting, with stronger overshooting at higher masses to match the observed  width of the main sequence band. They also note that the inclusion of stellar rotation does not ameliorate the problem of width of the modelled main sequence band.}

%%%%%%%%% TABLE %%%%%%%%%%%
\begin{deluxetable}{ccccc}
\tablecaption{The observed and predicted B/R \& Y/R ratios \label{tab:br_comparison}}
%\tablewidth{0pt}
%\tabletypesize{\scriptsize}
\tablehead{
\colhead{ratio} & \colhead{Observed} & 
\colhead{Observed$^{*}$} &\multicolumn{2}{c}{Predicted} 
\\ [-2ex] 
\colhead{} & \colhead{} & \colhead{} &
\colhead{constant $P(f)$} & \colhead{Gaussian $P(f)$}
}
%\colnumbers
\startdata
B/R & \phn 97/430=0.23 & 0.14 & 0.15\phn & 0.15\phn \\
Y/R & \phn 87/430=0.20 & 0.13 & 0.044 & 0.058 \\
(B+Y)/R & 184/430=0.42 & 0.26 & 0.19\phn & 0.21\phn
\enddata
\tablecomments{The observed supergiant candidates in the \citet{Neugent:2012aa} data are selected based on their LMC membership\deleted{ determination}. The blue, yellow, and red supergiants (B/Y/RSG) are then separated based on their effective temperature using the criteria described in section \ref{sec:obs_HR}. Furthermore, the candidates that have 4.0 $<$ \lgL $<$ 5.0 are selected to calculate the observed ratios listed in this table (see last row in Table \ref{tab:supergiants}). The column marked with asterix (*) shows a modified observed ratio if we include the 60\% missing dust-enshrouded RSGs as discussed in section \ref{subsec:dust-enshrouded_RSG} \deleted{\ref{sec:obs_BR}}. The predicted ratios are calculated using model time-scales for stars with masses in the range of 12--15 \msun , as explained in section \ref{sec:BR_calc}. $P(f)$ is a probability function of finding a star with a particular $f$ value for a given mass.}
\end{deluxetable}

\subsection{Pre-Supernova Radii and Final Masses}

As we have explored the evolution of massive stars starting with different initial masses and (exponential) overshoot factors, we find that the luminosity of a given star is highly dependent on the above variables at any given evolutionary stage (e.g. the Terminal Age Main Sequence (TAMS), the horizontal transition from the BSG to the RSG phase, the He-ignition stage, etc.). This is evident from our Figure \ref{fig:hrd} (or the Figure 1 of our Paper I). Depending on the mass and the overshoot factors, the stars spend different fractions of their post-TAMS lifetimes in either the RSG phase or the other hotter phases. The mass loss rates from the star is known to be sensitively dependent on both Luminosity and \Teff \ \citep[see for example figure 6 of ][ and the references listed therein]{Wagle:2020aa}. For a star of a given mass and an overshoot factor, the mass lost from it up to the core-collapse stage depends mainly on the fractional duration that the star spends in the RSG phase. It is in this phase that the mass loss rate is particularly high. As the total post-main-sequence lifetime, as well as, the fractional duration in the RSG phase depend on the initial mass and the overshoot factor, it is evident that the final mass of the star before core collapse would be different.

In Figure \ref{fig:mass_variation}, we show the final masses and the final radii of the stars as a function of the overshoot factor for different initial masses. For certain combination of masses and overshoot factors, the star spends long time in the hotter phases due to the blue loops. If the star lacks blue loops altogether, it would end up losing more mass. It will then end up with low final mass than it otherwise would, as seen for initial masses 13 \msun , 14 \msun , and 15 \msun \ star. The final radius tends to vary monotonically with the overshoot factor by nearly a factor of two for certain initial masses (12 \msun \ and 13 \msun).%However, at larger initial masses (14 \msun \ and 15 \msun) this monotonic trend of increasing radii is broken only at higher overshoot factors.

\added{The referee of this paper has  pointed out that \citet{Farrell:2020aa} derive the luminosity of the star at the
end of the core carbon burning as a function of the final helium core mass. In addition, since the effective temperature of the star during the red supergiant phase is nearly constant at such late stages \citep[typically a few hundred
to a thousand years before core collapse for a 13 \msun \ star - see Fig. 2 and 3 and Table 2 of our Paper I,][]{Wagle:2019aa}, this may imply that the photospheric radius of the star varies with the (helium) core mass. This is indeed the case as our Figure \ref{fig:mass_variation} shows.
Note however, that the relation in equation (3) of  \citet{Farrell:2020aa} was derived with a constant  overshoot parameters $f_{OV,core} = 0.016$ and $f_{OV,shell} = f_{OV,env}= 0.0174$ in the MIST implementation \citep{Choi:2016aa} with $\alpha_{MLT} = 1.82$ (similar to our  $\alpha_{MLT} = 2.0$), whereas here we explore the evolution of the stars for not only different initial masses $M$, but also for variable and
increasing overshoot factors $f$ with mass. It is well-known that a higher overshoot factor leads to a higher He core mass at the end stage for a given initial mass. Thus, while with varying
initial mass and overshoot factors, the core helium mass may be changing, their effects through the surface luminosity of the star may translate into a variable photospheric radius of the pre-SN star. We depict in Figure \ref{fig:mass_variation} the overall variation of the final radius of the star as a function of $(f, M)$. As pointed out by  \citet{Farrell:2020aa},
the helium core mass determines the mass of the compact remnant left behind after the supernova and also influences the nucleosynthetic and chemical yields. We describe in our Paper I the variation of the compactness parameter with overshoot factors for a 13 \msun \ star. We indicate in the next section how the different initial masses and overshooting strengths may cause another effect, namely the observable luminosity display of the
supernova, post-explosion.
}

\section{Discussion and Conclusion}\label{sec:conclusion}

In this work, we have presented models for progenitors of Type IIP supernovae, using nearby, well-observed SN 2013ej in the host galaxy M74 for an example. The metallicity of the neighboring H\textsc{ii} region 197 of SN 2013ej is determined to be 0.006 \citep{Cedres:2012aa}, which is similar to that of the LMC. We therefore investigate the evolution of the progenitor star in a low Z environment up to the pre-supernova stage using the Ledoux criterion for convection and near-standard mass loss rate predicted for hot and cool stars. In continuation of our previous papers in this series, we study the effects of convective overshoot on the model predicted B/R ratio in this paper. We try to determine how well the model properties reproduce the observed B/R ratio for the LMC. 

The luminosity of supergiant star is affected by both its initial mass, as well as the extent of overshooting. Thus, we explore a grid of masses and overshoot factors. We find that the key feature displayed by these models in the HR diagram is the blue loop (see our Paper I for an analysis.) There is a range of overshoot parameter ($f$) values that for a given intital mass exhibit the blue loop behavior. However, outside of this range, the blue loops disappear. In addition, the maximum and minimum values of $f$ that exhibit a blue loop follow a linear trend in mass (as seen in Fig. \ref{fig:hrd}). The blue loop is not exhibited in the models with initial masses higher than about 16 \msun . This is also seen in the figure, where the two lines (orange and blue) fitted to the maximum and the minimum $f$ values that undergo a blue loop intersect. The existence of the blue loop is in fact critical for accounting the observed number of BSG stars. The models that undergo a blue loop spend a substantial amount of time in their post-TAMS lifetime at relatively higher effective temperature regions. In contrast, the models with higher or lower $f$ values transit quickly to RSG branch and spend their entire post-core-helium-ignition lives in the RSG stage. 

The observations by \citet{Neugent:2012aa}, however, only extend to a maximum \Teff \ of about 10,000 K. Therefore, we subdivided the supergiant region into the blue, yellow, and red regions based on the effective temperature. 
%as follows. We refer to stars with 4.0 $\ge $\lgTeff $ >$ 3.875 as the BSG, 3.875 $\ge $ \lgTeff $ >$ 3.7 as the YSG, and 3.68 $\ge$ \lgTeff \ as the RSG candidates. 
We use these criterion for determining the B/R and Y/R ratios for both the observations as well as the model predictions. We have determined the B/R and Y/R ratios predicted by our \textit{non-rotating} stellar models in the mass range of 12--15 \msun , using the formulation explained in section \ref{sec:BR_calc}. We are able to match very well the observed B/R ratio in the LMC, especially if we include the 60\% missing RSGs that might be dust-obstructed (see section \ref{sec:obs_HR} and Table \ref{tab:br_comparison}, for comparison). However, the Y/R ratio is under-predicted by our models within a factor of $\sim$2 in comparison to the observations. If we group together the observed B+YSG stars then the predicted (B+Y)/R ratio comes within a good agreement within a few tens of percent. %Note however that the number of RSGs can be larger by a factor up to 1.6 due to some of these being shrouded in high extinction circumstellar dust shells (see section  2.1). If the number of RSGs are upto 60\% larger then, the $B/R$ and $Y/R$ observed ratios are correspondingly smaller and the model results are in nearer agreement with observed data. 
%\citet{Neugent:2012aa} predict well the relative number of YSG using Geneva models in the same range of effective temperatures as defined above. However, they do not attempt to predict the B/R ratio. The main difference in their work and this work is the YSG lifetimes predicted by their models. We attribute these differences to their choice of parameters used for the stellar evolution models. 
We would also like to note that we have chosen all of the B/Y/RSG stars present in the data within the chosen luminosity range for calculating the ratios. However, the spectral types are not determined for all of these stars, especially most of the RSG candidates. In fact, if we were to choose the spectroscopically identified candidates only for our calculations, then we would have about 85-95 \% of the yellow and blue supergiant candidates available to us from the data. However, we would account for only about 14\% of the total number of RSG candidates. Thus, selecting only the spectroscopically identidied candidates would skew the B/R and Y/R ratio to a higher side (a value $>$ 1). However, it would not be a correct comparison as many genuine RSG candidates would be neglected in such case. %On the other hand, in our current comparison we might be overestimating the observed RSG candidates that are single stars, much more significantly than the blue and the yellow supergiants. However, currently it is not possible to estimate how many RSG candidates are single stars without a proper spectroscopic determination.

\added{The presence of the blue loop in the models and the distribution of LMC supergiant stars in the HRD suggest that convective overshooting may be stronger for more massive stars.} In this paper, we have explored predictions for both mass and overshoot factors affecting the stellar luminosity and temperatures at different stages of its evolution through model calculations and used the lifetimes in specific segments of the HRD to compute the B/R or the Y/R ratios implied from these model calculations. Our attempts to match these predictions with observations of supergiants in the LMC are largely successful. Blue loops due to their long lifetimes of these phases are important elements in matching the model predictions to observations. The predictions of blue loops in the HRD for certain ($M, f$) ranges implies that the extent of the overshoot for which blue loops are found are bounded between two extreme $f$ values which may in turn be \replaced{varying}{intensifying} with the initial mass $M$ (see Figure \ref{fig:ov_range}).

\added{Stellar rotation affects the (computed) ratio of Blue and Yellow Supergiants with respect to Red Supergiants mainly through interior chemical mixing and thereby
altering the lifetimes of the stars in the different effective temperature ranges or spectral classes. The interplay of rotation and internal angular momentum transfer, metallicity and mass loss from the star 
in this context
%and their effects
%on the changes of the ratio of the stellar rotation speed to the critical speed of breakup as the massive star evolves 
was discussed by \citet{Maeder:2001aa} using the Geneva stellar evolution code. 
%The inclusion
%of rotation in stellar models alleviates the problem of the large number of RSGs observed in low metallicity galaxies, since the physical effects of rotation favor a redward
%evolution in the HR diagram. Rotation enhances the mass loss rates which in turn contributes to favor the formation of the red supergiants. 
Rotation induces chemical
mixing in the main sequence phase leading to a slight extension of the core (over its non-rotating counterpart), which in turn favors the redward motion during the He burning
phase, just like in convective overshooting. 
Rotational mixing during the main sequence phase is the key reason for the formation of red supergiants at low Z.
%increases the amount of helium near the H-burning shell and in turn favors a red location of the star in the HR diagram. On the other hand, enhanced (rotational) mixing during the He-burning phase may maintain a star in a blue location in the HRD, through the
%activation of an intermediate convective zone associated with the H-burning shell. 
Note however that the effect of rotation on the Blue to Red SG ratio
\citep[see Table 2 and Section 7 of][]{Maeder:2001aa} makes a substantial difference only for relatively high rotation speeds $v_{ini} = 300 \; \rm km \; s^{-1}$ and has the correct trend only for
initial masses M $>$ 15 \msun \ . For lower masses like 12 \msun \ , B/R in fact {\it increases} with rotation from the already high value (compared to observed numbers in low metallicity galaxies like the SMC) reported in their work.}

\added{For YSGs in the LMC, \citet{Neugent:2012aa} compute and compare the lifetimes for non rotating and strongly rotating (at 40\% rotation rate of the critical breakup speed) stars with the Geneva code for the initial mass-range of 12--40 \msun \ . Once again whether the YSG lifetimes with rotation are shorter or longer than the case with no rotation depends on the initial
mass of the star (see their Table 6). They make a mass dependent comparison of the relative number of YSGs in observations versus models after normalizing them to each other
in the 12--15 \msun \ range and claim that the data match the strongly rotational models better than the no-rotation models. Unfortunately since this normalization involves the critical
12--15 \msun \ range where the data and the models are most reliable, a more detailed comparison between the actual numbers seen in the LMC data and model predictions in
this range (with strong rotation) is not possible. On the other hand, a comparison of non-rotating models of YSGs normalized with respect to RSGs are explicitly possible both
in the data and our (non-rotating) models in the 12--15 \msun \ range and we show that our B/R ratio match the data very well, while the Y/R ratio is matched within a factor of $\sim 2$.} 

Final masses and radii of the presupernova star affect the post explosion dynamics and development of light curves and spectra  significantly. They control the optical, IR display of the type IIP supernovae through both the amount of hydrogen left on the star at explosion stage as well as the radius $R(0)$ at shock breakout phase (which depends on the final presupernova radius $R_{fin}$ itself). For example the luminosity of the supernova at the plateau phase can be written as \citep[see equation (64) of][]{Arnett:1980aa}: 
$$  L \propto  R(0) v_{sc}^2 $$
where $v_{sc}$ is a constant that sets the velocity scale \citep[see][equation (5)]{Arnett:1980aa} and may depend upon the kinetic energy and mass of the star and their distributions at the time of shock breakout. Thus, the convective overshoot factor of the \textit{pre-}supernova star is expected to affect significantly the luminosity and spectral properties in the \textit{post-}supernova stage. Note that the spectral evolution of the developing supernova including its line properties is controlled by $v_{sc}$. In a future paper, we shall explore in detail how the convective overshoot affects the model light curves and spectra of a supernova, through the mass, radii and velocity scales controlled by the energy of explosion. %and subject the model calculations to observed properties of real type IIP SNe (e.g. SN 2013ej). 
%A significant aspect of our model calculations is that different overshoot factors operating in stars of the same initial mass may lead to dramatically different pre-supernova stellar radii. In addition, stars with two different masses and overshoot factors may end up in a pre-supernova star with similar final masses but vastly different final radii (see Figure \ref{fig:mass_variation}) Differences in final masses prior to the supernova stage are also significant for different overshoot factors, due to the different lifetimes spent in the RSG phase where the mass loss rates are significant. Both these factors may affect the post-supernova optical/IR light curves at plateau phase of type IIP supernovae, since the initial radius $R(0)$ at shock breakout and the scale factor for post supernova expansion velocities $v_{sc}$ control the plateau luminosity while the hydrogen envelope mass (together with explosion energy) affects the plateau duration. We shall describe the effects on supernova light curves and spectral line evolution due to different treatments of convective overshoot at the pre-supernova phase in a future publication.

\section*{Acknowledgments}
We thank the directors and the staff of the Tata Institute of Fundamental Research (TIFR) and the Homi Bhabha Center for Science Education (HBCSE-TIFR) for access to their computational resources. This research was supported by a Raja Ramanna Fellowship of the Department of Atomic Energy (DAE), Govt. of India to Alak Ray and a DAE postdoctoral research associateship to Gururaj Wagle. The authors thank the anonymous referee for his/her constructive comments that nudged us to split off the work reported here from an earlier version of the original submission for Paper I. %GW thanks Rob Farmer for valuable feedback through private communication. 
\added{We also thank an anonymous referee of this paper for pointing out the \citet{Castro:2014aa} and \citet{Farrell:2020aa} papers. We thank Georges Meynet for discussions at the Conference on "Chemical Elements in the Universe: Origin and Evolution" in celebration of 150 Years of the Periodic Table at Bangalore, India.}
Adarsh Raghu thanks the NIUS program at HBCSE (TIFR). The authors thank NIUS participant Ajay Dev for his help in the selection of archival observations.
The authors acknowledge the use of NASA's Astrophysics Data System and the VizieR catalog access tool, CDS, Strasbourg, France.

\software{MESA r-10398 \citep{Paxton:2011aa,Paxton:2013aa,Paxton:2015aa,Paxton:2018aa}, Anaconda Spyder (Python 3.6)} 

%%%%%% END OF MAIN TEXT %%%%%%%%%

%%%% Bibliography %%%%%%%%
%\FloatBarrier
%\setlength{\bibsep}{0.0pt}
%\footnotesize
\bibliography{./my_bib}
%%%%%%%%%%%%%%%%%%%%

\end{document}